\documentclass[10pt, twocolumn, floatfix, superscriptaddress, aps, prx]{revtex4-2}

\usepackage{amsmath, amssymb, graphicx, hyperref, xcolor, colortbl, braket, cleveref, comment, mathtools}

\usepackage{pgfplots}
\usepgfplotslibrary{fillbetween,decorations.softclip,groupplots,external}
\pgfplotsset{compat=newest}
\pgfplotsset{plot coordinates/math parser=false}
\newlength\figureheight
\newlength\figurewidth

\usepackage{tikz}
\usepackage{tikz-3dplot}
\usetikzlibrary{3d, spy, external, matrix, plotmarks, pgfplots.groupplots, calc, decorations.pathreplacing, decorations.pathmorphing, arrows, decorations.markings, fit, backgrounds, arrows.meta, positioning, intersections}
\definecolor{myyellow}{RGB}{240,217,1}
\definecolor{mygreen}{RGB}{143,188,103}
\definecolor{myred}{RGB}{234,38,40}
\definecolor{myblue}{RGB}{53,101,167}
\definecolor{mygray}{RGB}{192,192,192}
\pgfdeclarelayer{bg}
\pgfsetlayers{bg,main}

\DeclareMathOperator{\tr}{tr}
\DeclareMathOperator{\avg}{avg}

\begin{document}

\title{Dynamic syndrome decoder in volume-law phases of hybrid quantum circuits}

\author{Dawid Paszko}
\affiliation{Department of Physics and Astronomy, University College London, Gower Street, London, WC1E 6BT, UK}

\author{Marcin Szyniszewski}
\affiliation{Department of Physics and Astronomy, University College London, Gower Street, London, WC1E 6BT, UK}
\affiliation{Department of Computer Science, University of Oxford, Parks Road, Oxford OX1 3QD, UK}

\author{Arijeet Pal}
\affiliation{Department of Physics and Astronomy, University College London, Gower Street, London, WC1E 6BT, UK}

\begin{abstract}
  Phases of matter with volume-law entanglement are frequently observed in quantum circuits and have numerous applications, ranging from deepening our understanding of quantum mechanics to advancements in quantum computing and cryptography. Their capacity to host entangled, complex quantum information is complemented by their ability to efficiently obscure it from quantum measurements through scrambling, reminiscent of quantum error-correction. However, the issue of initial-state decodability has primarily been studied in measurement-only models with area-law phases, which limit the entanglement of the encoded state. In this work, we introduce a class of Clifford circuits in one and two dimensions that feature a decodable volume law phase, allowing for information retrieval in logarithmic circuit depths. We present the \textit{Sign-Color Decoder} that tracks stabilizers revealing the initial state, akin to monitoring a dynamically-changing syndrome for error-correcting codes. We demonstrate this approach in scenarios where error locations are either known or unknown to the decoder, and we provide new insights about the relationship between the decodability transition and measurement-induced phase transition. We propose that this decodability transition is universal across various settings, including different circuit geometries. Our findings pave the way for using volume law states as encoders with mid-circuit measurements, opening potential applications in quantum error correction and quantum cryptography.
\end{abstract}

\maketitle

\section{Introduction}

Closed quantum many-body systems are known to generically thermalize under unitary dynamics~\cite{Deutsch1991, Srednicki1994, DAlessio2016, Borgonovi2016}. This implies that the local information in the initial state is transformed into non-local operators delocalized throughout the system, making it inaccessible to local measurements. However, some exceptional nonintegrable systems defy the eigenstate thermalization hypothesis, such as those containing quantum many-body scars~\cite{Moudgalya2018, Moudgalya2018scars, Turner2018nat, Turner2018, Lin2019, Schecter2019, Iadecola2020}, or many-body localization~\cite{Basko2006, Gornyi2005, Pal2010, Nandkishore2015, Abanin2019}. 
These systems preserve quantum information for long times~\cite{Bernien2017}, and can dynamically remain area-law entangled for quantum scars or grow exponentially slowly towards volume-law entangled states in a many-body localized system.  Conversely, fast generation of volume-law states in hybrid quantum circuits through scrambling prevents easy extraction of useful information~\cite{Choi_QECC2020}. Overcoming this challenge is crucial for using many-body states in quantum error correction~\cite{dennis2002topological, Kitaev2003, Bacon2017, Berthusen2025, Bravyi2025, delfosse2023spacetime}. A large class of area-law states has been recently shown in experiments to be promising for encoding and decoding quantum information in the logical qubit~\cite{bluvstein2024logical, google2025QEC, brock2025quantum, sommers2025FaultTolerance}. 

There are fundamental questions associated with the complexity of controlling highly-entangled states and leveraging them for quantum information processing~\cite{Li2021, yi2024complexityAQEC, Paszko2024SPT, Sang2024MixedQEC, Sommers2025}. On one hand, certain kinds of entanglement can \textit{screen} logical information from noise, it is certainly challenging to decode information and control entanglement without the state losing its error-correcting properties~\cite{anwar2014fastdecoders, Delfosse2020Decoder, delfosse2021LTDecoder, Li2023}. 
Recent studies have demonstrated a dynamical transition between volume-law and area-law phases by interspersing unitary evolution with quantum measurements, resulting in a measurement-induced entanglement transition (MIET)~\cite{Chan2019, Skinner2019, Li2018, Li2019, Szyniszewski2019}. Interestingly, these toy models exhibit extremely rich physics, relevant to the fundamental properties of entanglement spreading~\cite{Nahum2017, Cao2019, Chan2019, Skinner2019, Li2018, Li2019, Boorman2022}, mapping between quantum circuits and (classical and quantum) statistical physics models~\cite{Bao2020, Jian2020, Li2021, LiVasseur2021, Chahine2023, Fava2023}, and leading to new protocols for quantum state-preparation and robust information protection~\cite{Roy2020, Muller2022, Iadecola2023, Buchhold2022, Sierant2023, ODea2024, Sierant2023pol, Friedman2023, Piroli2023, Ravindranath2023, LeMaire2024, Qian2024, Angelidi2023}.
Given the simplicity of this setting, it provides an ideal framework for studying how information can be encoded, scrambled, and decoded using a variety of entangled phases of matter.

\begin{figure}[b]
    \centering
    \includegraphics[width=\columnwidth]{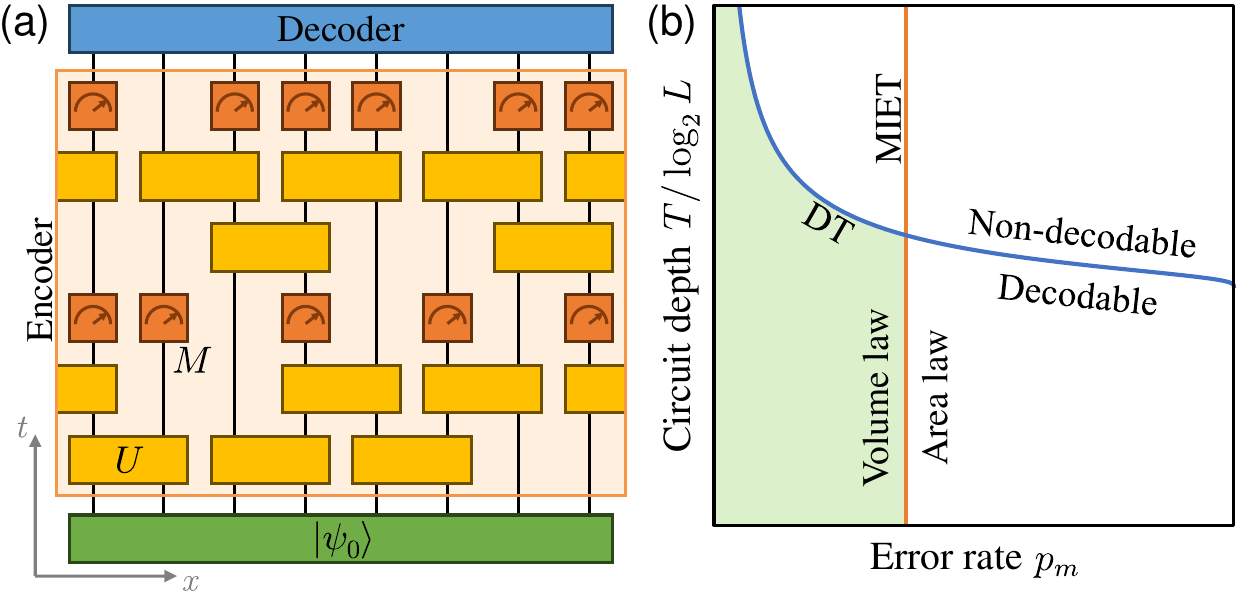}
    \caption{(a)~Quantum circuit considered in this work. Initial state $|\psi_0\rangle$ is evolved stochastically using unitaries $U$ and measurements $M$. At circuit depth $T$, the decoder attempts to decipher the initial state. (b)~Schematic phase diagram of the circuit as a function of measurement rate $p_m$ and the log coefficient of the circuit depth $T / \log_2 L$, showing a measurement-induced entanglement transition (MIET) and a decodability transition (DT). The green region is the decodable volume law phase.}
    \label{fig:circuit}
\end{figure}

Decoding of dynamical phases in MIET is also known under the name of learnability, and encompasses three major types of protocols: reconstructing the initial state~\cite{Li2019, Barratt2022, Li2023, Roser2023, Ippoliti2024}, determining a global property of the initial state (such as charge~\cite{Agrawal2022, Agrawal2024, Barratt2022, Barratt2022field, Feng2025sharp, Gopalakrishnan2025} or spin~\cite{Majidy2023, Feng2025sharp, Gopalakrishnan2025}), or identifying the dynamical phase of the system~\cite{Gullans2020dyn, Gullans2020, Yoshida2021, Dehghani2023, Hoke2023}. In this work, we mostly focus on the first case, where the initial state is encoded via a monitored quantum circuit, and decoded using either the measurements at the end of the circuit, the record of mid-circuit measurements, or the positions of measurements and gates. While the existing knowledge of decoding has primarily focused on area-law phases of measurement-only models~\cite{Nahum2020, Ippoliti2021, Lavasani2020, Sang2020, Lavasani2020topo, Lang2020, VanRegemortel2021}, such as the projective transverse field Ising model~\cite{Lang2020, Li2023, Roser2023}, 
we aim to study whether quantum information can be encoded within the volume law phase -- a question that remains largely unexplored in this context. For erasure errors, decoding is known to exhibit polynomial complexity even for random stabilizer codes, however is expected to be hard for general class of errors~\cite{Gullans2021_QCoding, Nelson2025}.
Developing an encoding-decoding protocol that utilizes volume-law states could unlock many applications, particularly in quantum cryptography by securing complex entangled quantum information through controlled scrambling, or in quantum error correction by improving the code distance.

In this paper, we present, to our knowledge, the first example of a protocol that enables quantum state retrieval within the volume-law phase. We introduce the \textit{Sign-Color decoder}, which tracks stabilizers correlated with the information of the initial state, allowing the syndrome operators to dynamically evolve with the depth of the circuit. Our analytical and numerical results reveal that this protocol enables decoding of the initial state in logarithmic circuit depths, regardless of whether the error locations are known or unknown to the decoder. Moreover, we uncover fundamental insights into the critical properties of the decodability transition itself, highlighting its high degree of universality across different circuit geometries. By harnessing the volume-law generating circuits with mid-circuit measurements as encoders, our results provide a foundation for their future applications in quantum error correction and quantum cryptography.

This manuscript is organized as follows. Sec.~\ref{sec:protocol} introduces the monitored circuit, which we show to exhibit a MIET, and hence host the volume and the area law. In Sec.~\ref{sec:decoding_located}, we introduce the Sign-Color decoder and show that the decoding in the volume-law phase is possible in various one- and two-dimensional geometries, when the circuit depth is at most logarithmic in system size (see Fig.~\ref{fig:circuit}) and error locations are known to the decoder. We also provide a physical picture based on a stochastic mean-field model for the finite-size scaling of the time scales of the decodability transition. Furthermore, even when the information of the error locations is erased, we still find a decodable volume-law phase in Sec.~\ref{sec:decoding_unlocated}. Finally, we discuss the results and the future outlook in Sec.~\ref{sec:discussion}.

\section{Hybrid quantum circuits and entanglement transition}
\label{sec:protocol}

We consider the following quantum circuit, where at every time step there are two-site unitary gates $U$ followed by measurements $M$ of the Pauli $X$ operator. The unitaries are of the form
\begin{equation}
    U_{ij} = \exp\left[- i \frac{\pi}{4} (X_i X_j + Z_i Z_j)\right],
    \label{eq:unitary_gate}
\end{equation}
where $X$ and $Z$ are Pauli matrices; $U_{ij}$ are Clifford gates. The circuit is stochastic, where each quantum operation has a finite probability of occurrence. The gates $U$ are applied in a brickwork fashion with probability $p_u$, as shown in Fig.~\ref{fig:circuit}(a), while the measurements are applied on all sites with probability $p_m$. Intuitively, we will consider the gates to act as encoders of information, while the measurements can be viewed as errors in the encoding process.

\begin{figure}
    \centering
    \includegraphics[width=\columnwidth]{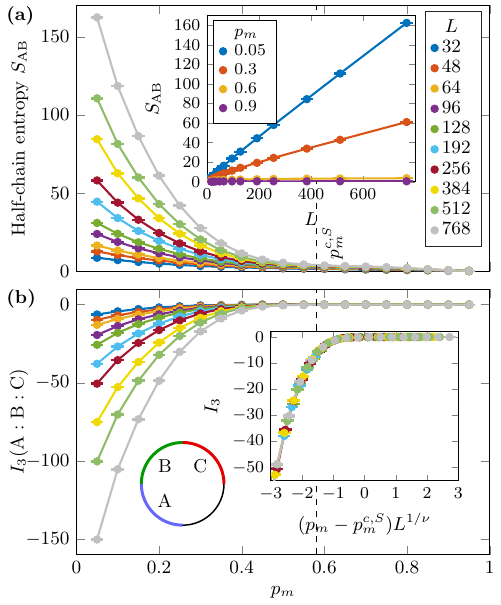}
    \caption{(a)~Half-chain entanglement entropy $S_\text{AB}$ as a function of the measurement frequency $p_m$ for $p_u=0.9$ and different values of system size $L$, at circuit depth $T=16 L = O(L)$. The inset shows entropy vs system size dependence. (b)~Tripartite mutual information $I_3(\text{A}:\text{B}:\text{C})$. The legend in panel (a) applies in (b). The inset plot shows a data collapse at $p_m^{c,S} = 0.57(4)$ and $\nu = 3.3(4)$. The inset circular diagram shows how the system is divided into four regions, each of length $L/4$. The dashed line shows the transition point.}
    \label{fig:1d_miet}
\end{figure}

As we show below, this circuit exhibits a measurement-induced phase transition from the volume-law to the area-law phase when $p_m$ is increased or $p_u$ is decreased. It is analogous to a circuit studied earlier by two of the authors for a state-preparation protocol~\cite{Angelidi2023} where the measurements are applied on alternate sites, which also exhibits this transition.
In order to characterize the measurement-induced transition in our hybrid circuit, we will use the von Neumann entanglement entropy between subsystem R and its complement $\bar{\text{R}}$, defined as $S_\text{R} = -\rho_\text{R} \ln \rho_\text{R}$, where $\rho_\text{R} = \tr_{\bar{\text{R}}} (\ket{\psi}\bra{\psi})$ is the reduced density matrix of R. Generally, we will be interested in half-chain entanglement entropy. To pinpoint the transition threshold and its critical properties, we will also use the tripartite mutual information~\cite{Zabalo2020},
\begin{align}
    I_3(\text{A}:\text{B}:\text{C}) &= S_\text{A} + S_\text{B} + S_\text{C}\nonumber\\
    &\quad- S_\text{AB} - S_\text{AC} - S_\text{BC} + S_\text{ABC},
    \label{eq:tmi}
\end{align}
defined for three adjacent regions A, B, and C. Specifically, we will use continuous regions of size $L/4$, as shown in the inset of Fig.~\ref{fig:1d_miet}(b), which cancels the finite size effects due to the boundary contributions. 

The circuit is initialized in the all-minus state $\ket{\psi_0} = \ket{-}^{\otimes L}$ and is evolved until circuit depths $T=16 L = O(L)$. The results for half-chain entanglement entropy $S_\text{AB}$ and tripartite mutual information $I_3$ for $p_u=0.9$ are shown in Fig.~\ref{fig:1d_miet}. The entropy shows a clear change in behavior between growing linearly with the system size (volume law) for small values of $p_m$ and saturating to an $O(1)$ value (area law) at large values of $p_m$, signifying an entanglement transition. We collapse the data for $I_3$ for large system sizes $L \ge 192$ using finite-size scaling ansatz~\cite{Zabalo2020}
\begin{equation}
    I_3 = G[(p_m - p_m^{c,S}) L^{1/\nu}],
\end{equation}
where $p_m^{c,S}$ is the critical measurement probability for the entanglement transition, and $\nu$ is the critical exponent of the correlation length, and $G[\cdot]$ is a universal one-parameter function. See Appendix~\ref{app:fss} for more details. The resulting data collapse is of excellent quality, as shown in the inset of Fig.~\ref{fig:1d_miet}(b), with critical parameters $p_m^{c,S} = 0.57(4), \nu = 3.3(4)$.

The quality of the collapse suggests that the transition is of second order, similarly to other volume-to-area-law entanglement transitions found in monitored systems. Due to this similarity, we also believe that the transition becomes apparent at circuit depths of $T\sim \ln L$ (see the minimal cut picture~\cite{Skinner2019, Lunt2021}). By comparing the collapses for different values of $p_u \in \{0.3, 0.6, 0.9\}$ (see Appendix~\ref{app:supp_res}), we find that the critical exponent is approximately independent of $p_u$, showcasing the universality of this transition across the entire parameter regime.

\section{Decodability transition}
\label{sec:decoding_located}

We now formally define the decoding problem for this class of circuits. The unitary gates act as encoders that transform the initial state into a highly entangled complex state, which provides immunity to errors. On the other hand, local measurements mimic the effect of errors. The strategy is to encode the initial information into a dynamically evolving state in the volume-law phase and decode the initial state at the end of the evolution for a specific circuit depth $T$. As a result, the syndrome operator for the decoding protocol is not static, and depends on the specific realization of the dynamics, i.e., location of encoders and errors. We first demonstrate that efficient decoding is indeed achievable in this setting, in a simplified case, where the locations of the measurements are known (KL) is known to the decoder, but the measurement outcomes are not. Similarly, the initial logical state is assumed to contain a single bit of classical information.  Using the repetition code as an example, we initialize the state in either the all-plus state $|+\rangle^{\otimes L}$ or the all-minus state $|-\rangle^{\otimes L}$. In Sec.~\ref{sec:init_states}, we will show that our analysis can be extended to other sets of initial states encoding a finite number of logical bits in long-range entangled states. We propose a protocol to decode the initial state from the available information, the locations of errors and the measurement outcome of the syndrome measurement at the end of the circuit. The more realistic scenario with unknown locations (UL) of errors will be investigated in Sec.~\ref{sec:decoding_unlocated}.

\subsection{Sign-Color decoder and dynamic syndrome}

In this subsection, we introduce the primary decoding protocol introduced in this work, which we refer to as the Sign-Color decoder (SCD). The protocol relies on the state remaining a stabilizer state during the circuit evolution. SCD keeps track of state stabilizers that reveal the initial state by labeling each stabilizer with a color. As we will see, from the quantum error correction perspective, the coloring is equivalent to finding a syndrome that captures all available information about errors in the system -- this continuously evolving syndrome changes at each time-step, hence the name \textit{dynamic syndrome}. The outcome of the final syndrome measurement allows us to decode the initial state even in the presence of errors.

The evolution of a stabilizer state under Clifford gates and measurements of Pauli strings is classically simulable in polynomial time~\cite{Gottesman1996, Gottesman1998, Nielsen2010}. Here, we use the Aaronson-Gottesmann algorithm~\cite{Aaronson2004}, which represents the state as a tableau of its stabilizer generators (and destabilizers, which are used to speed up the performance of measurements). When we fix the circuit structure, i.e., the positions of gates and measurements, this corresponds to fixing the stabilizer generators of the final state, up to their signs. There are three distinct possibilities for how the sign of a stabilizer is related to the initial state. \textbf{(1)}~The sign is uncorrelated with the initial state, but also does not depend on measurement outcomes, therefore \textit{trivial}.  An example of such a stabilizer is a global symmetry of the state, $\prod_i X_i$, which has a trivial sign of $+1$ that is unaffected by unitary evolution and measurements. \textbf{(2)}~The sign is \textit{correlated} with the initial state, e.g.\@ a sign of $+1$ is associated with the initial state being the all-plus state $|+\rangle^{\otimes L}$, while $-1$ corresponds to $|-\rangle^{\otimes L}$.
\textbf{(3)}~The sign is \textit{randomized} by a measurement, and becomes uncorrelated with the initial state and only dependent on the measurement record.
We associate a color with each of the three options for the \text{sign} of a stabilizer generator for a time-evolved state, and their list of all stabilizer generators with a coloring of the state.

Hence, the decoding problem is equivalent to predicting the correlations between the signs of the stabilizer generators and the initial state. The coloring pattern of the state ought to contain a correlated sign for the state to be decodable. However, there is a property of the tableau, that renders extracting the correlations particularly difficult -- the stabilizer generators are not unique, and one can multiply one generator through another, creating a different tableau that corresponds to the same state. This non-uniqueness implies that correlations with the initial state can be hidden in collections of stabilizer signs, where they are required to be mutually multiplied to extract the correlated stabilizer. 
Within the coloring picture, this can be viewed as non-trivial color mixing, leading to exponentially many distinct stabilizer sign colorings. Naively, this would require tracking all possible $2^L$ stabilizers of the state in order to determine whether it is decodable conclusively.

\begin{figure}
    \centering
    \includegraphics[width=\columnwidth]{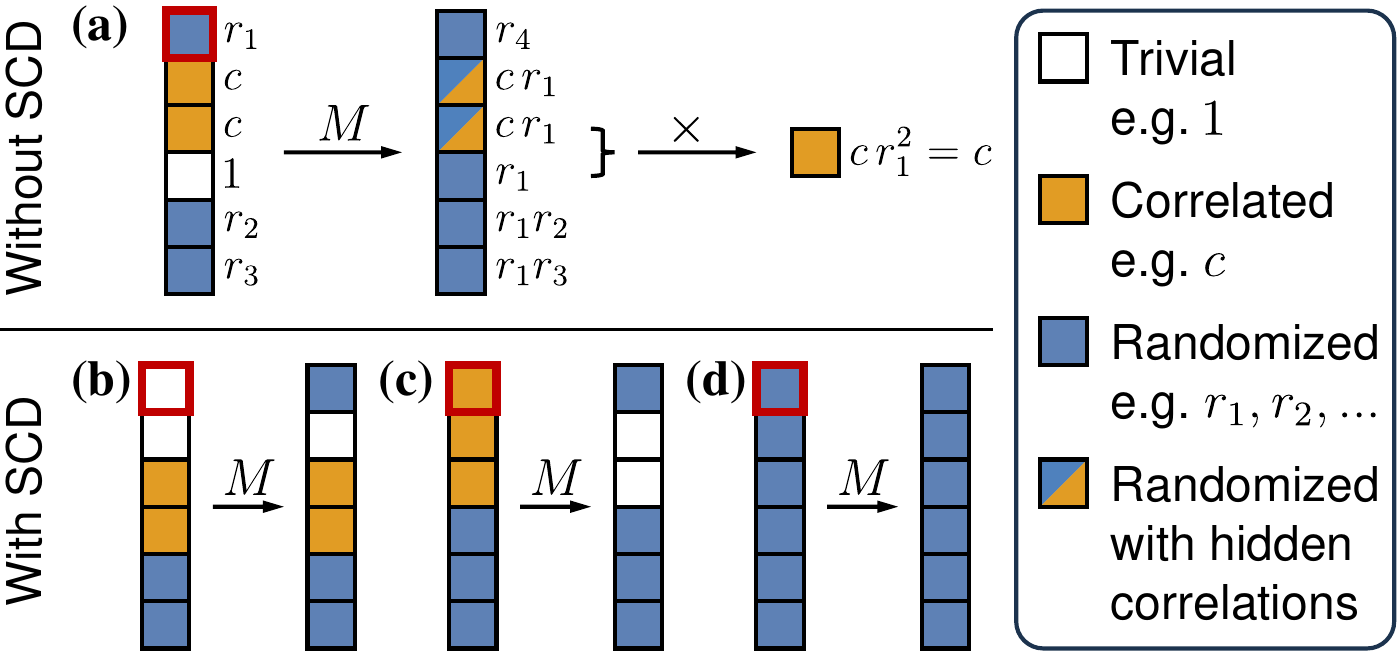}
    \caption{Examples of how sign colors change in the list of stabilizer generators that anticommute with the measurement operator $M$ during a measurement process. The selected sign is marked in dark red.
    (a)~Without the Sign-Color decoder (SCD), hidden correlations can arise. For example, we show how a randomized sign $r_1$ is multiplied by a correlated sign $c$ and a trivial sign $1$, allowing for the decoding only if one considers the full list of $2^L$ stabilizers. With SCD, the signs are sorted to prevent hidden correlations. Examples for the selected sign to be (a)~trivial, (b)~correlated with the initial state, and (c)~randomized.}
    \label{fig:signs_explanation}
\end{figure}

The color mixing is intimately related to how the generators change during the application of a quantum measurement. Note that the application of gates does not change the coloring of the signs, as it does not mix the generators. However, during a measurement, according to the Aaronson-Gottesmann algorithm, one of the anti-commuting generators with the measurement operator $M$ is arbitrarily chosen to be multiplied with the others in this set. Following which, the chosen anticommuting generator is replaced by the measurement operator $M$ with a \textit{randomized} sign.
The color mixing problem arises when the chosen stabilizer has a randomized sign. As illustrated in Fig.~\ref{fig:signs_explanation}(a), the information about the initial state is still available in the tableau, but it is now hidden under signs that appear randomized. As a consequence, decoding after an extensive number of measurements requires searching through the full space of stabilizers.

The SCD implements an algorithm that minimizes color mixing, such that the coloring state preserves the initial distinction between the trivial, correlated, and randomized signs as long as possible, so that no hidden correlations arise due to measurements. This ensures that the decoding can be performed efficiently in polynomial time by keeping track of $L$ stabilizer generators. During the measurement, we choose the anticommuting generator according to the priority of sign-colors: (1)~trivial, (2)~correlated, (3)~random. Then the only color changes are as follows. Multiplying any stabilizer sign by a trivial sign does not change its correlation with the initial state [see Fig.~\ref{fig:signs_explanation}(b)]. If no trivial signs are left in the list of anticommuting generators, we use a correlated sign, which turns other correlated signs into trivial ones, and leaves randomized signs as randomized [Fig.~\ref{fig:signs_explanation}(c)]. And if only randomized signs are left in the list, any multiplication does not change their nature [Fig.~\ref{fig:signs_explanation}(d)].

To summarize, the Sign-Color decoder tracks whether the stabilizer signs are trivial, correlated with the initial state, or random. The existence of a correlated sign in the state coloring permits decodability without changing the computational complexity. The coloring of a state serves as a \textit{dynamic error syndrome} which provides classical information associated with the errors in the circuit. Conventionally, the syndrome consists of parity measurement outcomes, which allow for locating and correcting the error. In an analogous manner, the dynamic syndrome reveals the history of errors by identifying the stabilizers impacted by the errors, and allows us to correct the state. The state is corrected by performing a measurement of the stabilizer generator with a correlated sign color that reveals the initial state. Having the protocol for exact decoding, we explore the threshold for decodability, and the time scales associated with it.

\subsection{Stochastic model for the decodability threshold}
\label{sec:simple_random}

\begin{figure}[t]
    \centering
    \includegraphics[width=\columnwidth]{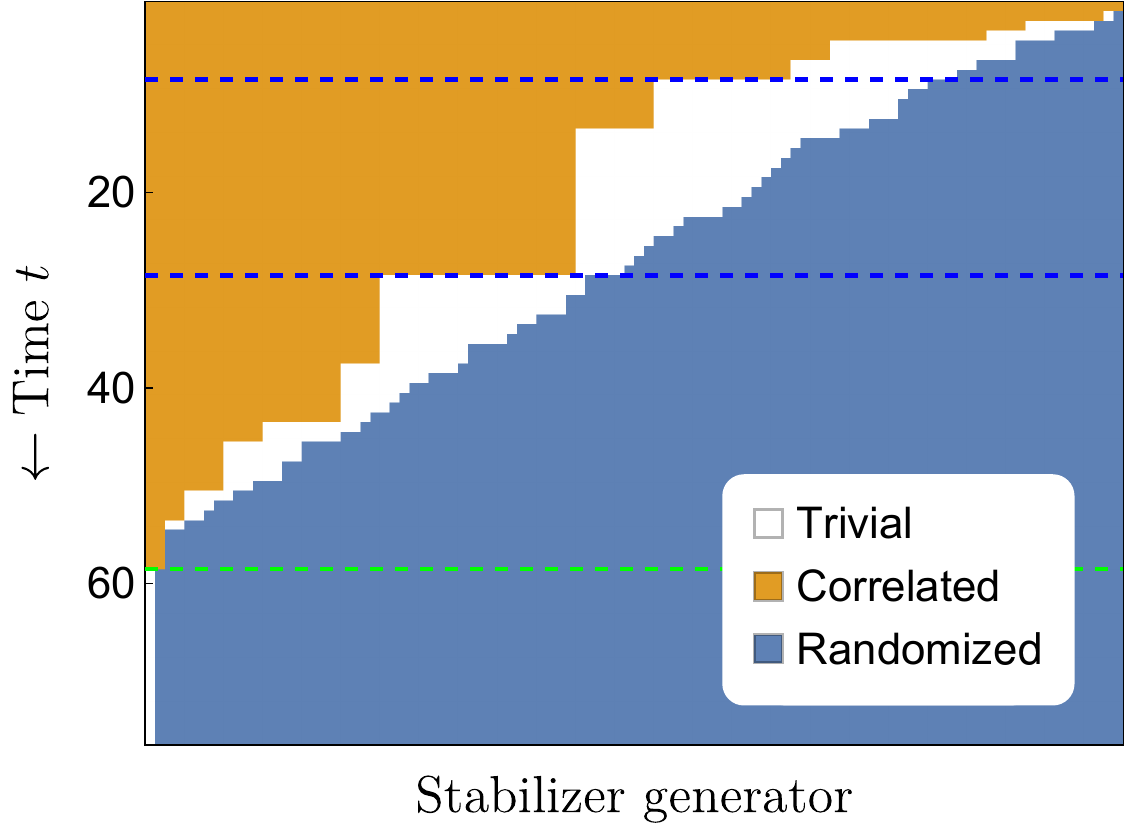}
    \caption{Example plot showing changes of the stabilizer signs in time. $L = 100, p_u = 0.9, p_m = 0.02$. Some measurements have a substantial impact on the signs (blue dashed lines). At time $T_r$ (dashed green line), all stabilizer signs become independent of the initial state. At long times, all stabilizer signs are randomized, except one: the global symmetry of $\prod_i X_i$, which has a trivial sign of $+1$.}
    \label{fig:history}
\end{figure}

In this section, we develop a stochastic theory (akin to a mean-field theory) of the circuit depth $T_r$, beyond which the state is no longer decodable, i.e.\@ all correlations with the initial state disappear, and generator signs become trivial or randomized. An example is shown in Fig.~\ref{fig:history}, where we plot the signs as a function of time for a single random realization of the circuit. The sign-colors correlated with the initial state disappear at time $T_r$, marked with a dashed green line. The signs seem to change randomly, with two forms of qualitative behavior: the number of correlated signs never increases, while the number of randomized signs never decreases. We also note that some measurements seem to impact many signs in a single time step, particularly those marked with dashed blue lines. This happens when no trivial anticommuting signs exist during a measurement operation. At long times, all signs are randomized, except one -- corresponding to the global symmetry of $\prod_i X_i$ with a trivial sign of $+1$.

The random process of transforming the sign structure of stabilizers containing predominantly correlated signs to randomized ones can be captured by a stochastic model, which we will use to estimate the behavior of $\bar T_r$. Consider a list of $L$ stabilizer signs. At each time step, assume that every sign independently has a probability $\mathcal{P}$ of becoming randomized, and does not revert back to being correlated at a later time.  Hence, the distribution of randomization time for a sign is expressed as a geometric distribution with probability density function (PDF) $f(t)$ and cumulative distribution function (CDF) $F(t)$,
\begin{equation}
    f(t) = (1-\mathcal{P})^t \mathcal{P},\quad \text{and} \quad
    F(t) = 1-(1-\mathcal{P})^{t+1},
\end{equation}
respectively. Assuming that the stabilizer signs are independent from each other, the distribution for the depth at which the entire system becomes uncorrelated is given by the $L$-th order statistic, with a PDF
\begin{equation}
    f_\text{os}(t) = (1-(1-\mathcal{P})^{t+1})^L - (1-(1-\mathcal{P})^t)^L.
    \label{eq:PDFsm}
\end{equation}
The mean circuit depth is then given by $\bar T_r = \sum_{t=0}^\infty t f_\text{os}(t)$, which on approximating the sum as an integral, can be evaluated to be
\begin{equation}
    \bar T_r \approx \frac{H_L}{\ln[1/(1-\mathcal{P})]} - \frac{1}{2} \approx \frac{\ln L + \gamma}{\ln[1/(1-\mathcal{P})]} - \frac{1}{2},
    \label{eq:Tr}
\end{equation}
where $H_L$ is the $L$-th harmonic number, $H_L=\sum_{i=1}^L 1/i$, and $\gamma$ is the Euler-Mascheroni constant. Hence, this simplifying approximation predicts that the mean circuit depth $\bar T_r$ should scale logarithmically with the system size.

\begin{figure}
    \centering
    \includegraphics[width=\columnwidth]{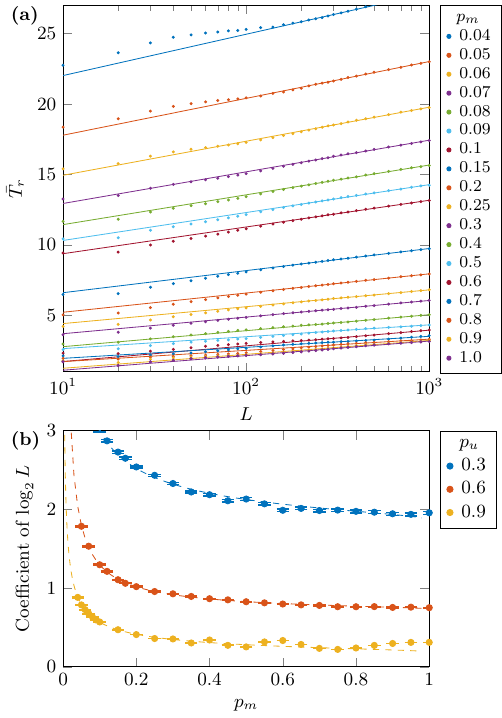}
    \caption{(a)~Mean circuit depth $\bar T_r$ after which the final state is uncorrelated with the initial state, for $p_u = 0.9$, as a function of the system size $L$. Solid lines are fits for $L \ge 220$. (b)~Coefficient of the $\log_2 L$ term for a fit to large system sizes. The dashed lines are fits to $c - a / \ln(1 - p_m^\beta)$.}
    \label{fig:mean_depth_1D}
\end{figure}

Our numerical results, shown in Fig.~\ref{fig:mean_depth_1D}(a), confirm that this prediction describes the mean circuit depth quite accurately. The logarithmic scaling of the depth can be seen across all values of $p_m$, which all fit well to an ansatz $\bar T_r = a \log_2 L + b$. However, at low $p_m$ finite size effects become increasingly more pronounced, at some point reaching the available system sizes. This means that extracting the coefficient of the $\log_2 L$ term is only accurate when $p_m$ is sufficiently larger than $0$.

In the limit $p_m \to 1$, we find that the coefficient of the logarithmic term converges to a finite value [see Fig.~\ref{fig:mean_depth_1D}(b)]. Measurements are always performed on all sites, and the stabilizer generators are of the form $\pm X_i$ with their signs initialized to be fully correlated. During the evolution, although the form of the generators after the measurement layer is fixed, their signs may change from correlated to either randomized or trivial. This random process leads to an equivalent calculation to that from Eq.~\eqref{eq:Tr}, albeit for a different $\mathcal P$. This then leads to a log behavior, explaining why the log coefficient assumes a finite value in the $p_m \to 1$ limit.
We plot the extracted log coefficient as a function of $p_m$ for a few values of $p_u$ in Fig.~\ref{fig:mean_depth_1D}(b), and find that it follows an approximate form of $c - a / \ln(1 - p_m^\beta)$ (see the dashed lines in the figure). This mimics the estimate from Eq.~\eqref{eq:Tr} (with $\mathcal{P} = p_m^\beta$ being some function of $p_m$).

In summary, we have successfully developed a stochastic framework that effectively captures the qualitative features of decodability in the quantum circuit. This description is akin to a mean-field theory, as we simplify the degrees of freedom within the quantum circuit by neglecting the quantum correlations and, in turn, dimensionality. In fact, this independence of certain model properties of dimensionality is a generic feature of self-consistent field theory models, and motivates further numerical investigations -- which we undertake in the following sections. The true mean-field description of the circuit should account for the temporal evolution of stabilizers and the spatial correlations of their signs. It may be insightful to study a similar decoding problem on a tree geometry~\cite{Lopez-Piqueres2020, Nahum2021, Feng2023, Ravindranath2025, Kim2025, Feng2025, Sommers2025, yadavalli2023noisyQT}, where the mean-field treatment becomes more tractable for measurement-induced entanglement transitions.

\subsection{Emergence of the decodability transition}
\label{sec:emergence_of_decodability}

\begin{figure}[t]
    \centering
    \includegraphics[width=\columnwidth]{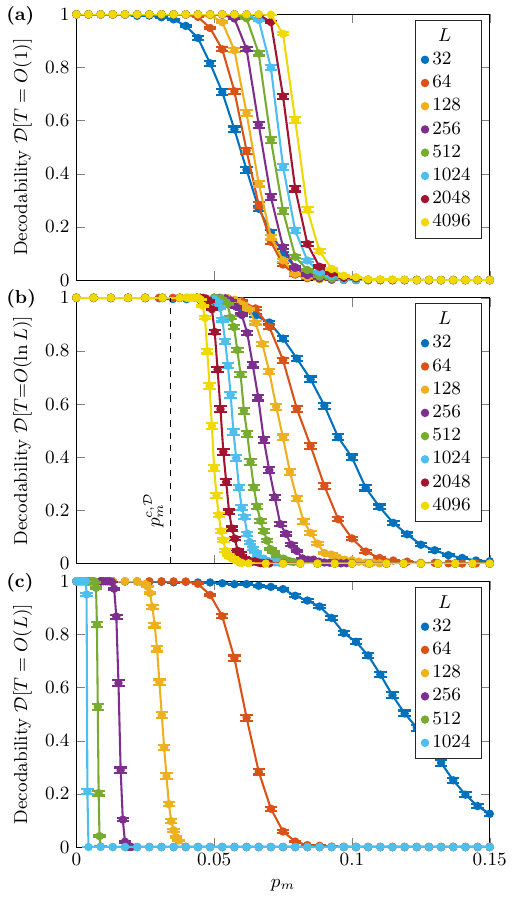}
    \caption{Decodability $\mathcal D[T]$ as a function of measurement frequency $p_m$ for $p_u=0.9$ and different system sizes $L$. The circuit depths are (a)~$T=16=O(1)$, (b)~$T=2\log_2 L=O(\ln L)$, and (c)~$T=L/4=O(L)$. The dashed line in panel (b) shows the decodability transition at the critical value of $p_m^{c,\mathcal{D}}$.}
    \label{fig:decodability_1D}
\end{figure}

We now turn our attention to calculations of decodability $\mathcal D [T]$, i.e., a probability that a random realization of the circuit of depth $T$ is decodable. Note that $\mathcal D[T]$ is defined in terms of the cumulative distribution function of $T_r$, as $\mathcal{D} [T] = 1 - \mathrm{CDF}(T_r = T)$. In this section, we mainly focus on $p_u = 0.9$ and varying $p_m$. Our numerical results are shown in Fig.~\ref{fig:decodability_1D} for three different scalings of the circuit depths: constant depths $T = O(1)$, log depths $T = O(\ln L)$, and depths extensive in the system size $T = O(L)$.

Since the system is decodable until approximately logarithmic depths, we expect the following behaviors. At depths $T = O(1)$, the system should always be decodable in the thermodynamic limit. Indeed, in panel (a) of Fig.~\ref{fig:decodability_1D}, we can see that the threshold for decodability slowly moves towards larger values of $p_m$ when $L$ is increased. Secondly, at extensive depths $T = O(L)$, we expect the system to never be decodable in an infinite system. Fig.~\ref{fig:decodability_1D}(c) showcases this clearly, with thresholds near $p_m = 0$ for large system sizes.

The most interesting situation happens at log depths of $T = O(\ln L)$. In this scenario, there exists a critical value $p_m^{c,\mathcal{D}}$, below which the system is always decodable, and above which it never is. This can be clearly seen in Fig.~\ref{fig:decodability_1D}(b), where the decodability thresholds drift towards $p_m^{c,\mathcal{D}} \approx 0.03$. The specific value of the critical point depends on the desired behavior of the circuit depths $T$, i.e., the coefficient of the logarithm. This can be read off from Fig.~\ref{fig:mean_depth_1D}(b); for example, when $T = \log_2 L$, the threshold can be estimated to be $p_m^{c,\mathcal{D}} \approx 0.01$. Ultimately, this leads to the phase diagram shown in Fig.~\ref{fig:circuit}(b). Note that when the coefficient of log is below some critical value, the initial state is always decodable for any value of $p_m$.

We expect the decodability transition to be a second-order phase transition. This is due to the similarity to the stochastic mean-field model from Eq.~\eqref{eq:PDFsm}, where it exhibits a continuous decodability transition (see Appendix~\ref{app:simple_random}). Interestingly, in the stochastic model, the universal critical exponent of the correlation length does not depend on the coefficient of the $\log_2 L$.
In order to extract the critical properties of the decodability transition in the stochastic circuit, we attempt a data collapse of the form
\begin{equation}
    \mathcal{D} = G[ (p_m - p_m^{c,\mathcal{D}}) \, l^{1/\nu} ],
\end{equation}
where $G[\cdot]$ is a universal function, $\nu$ is the critical exponent of the correlation length, and $l$ is the shortest lengthscale in the system, which in our case is the circuit depth $T \sim \ln L$.

\begin{figure}[t]         
    \includegraphics[width=\columnwidth]{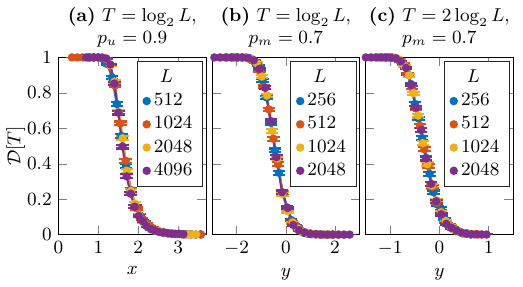}
    \caption{Data collapse for decodability $\mathcal{D}[T]$ for the following circuit depths and parameters: (a)~$T = \log_2 L$, $p_u=0.9$ (b)~$T = \log_2 L$, $p_m=0.7$, and (c)~$T = 2 \log_2 L$, $p_m=0.7$. The rescaled parameters are defined by: $x=(p_m - p_m^{c,\mathcal{D}}) (\log_2 L)^{1/\nu}$ and $y=(p_u - p_u^{c,\mathcal{D}}) (\log_2 L)^{1/\nu}$. The critical parameters obtained are listed in Appendix~\ref{app:supp_res} along with other data collapses.}
    \label{fig:1d_collapse}
\end{figure}

We perform data collapses at various circuit depths and parameters, and show some of them in Fig.~\ref{fig:1d_collapse}.
The data collapses are performed only for the largest available system sizes.
First, Figs.~\ref{fig:1d_collapse}(a) and (b) both display the transition for depth $T=\log_2 L$ but $p_u$ is fixed and $p_m$ is varied in panel (a), while $p_m$ is fixed and $p_u$ is varied in (b). The collapse results in $\nu = 0.74(20)$ and $\nu = 1.00(14)$, respectively. This suggests that there exists a transition line cutting across the $(p_u,p_m)$ space. The full phase diagram including these cuts is drawn in Fig.~\ref{fig:1d_phase_diagram}. Second, Figs.~\ref{fig:1d_collapse}(b) and (c) compare two different circuit depths $T=\log_2 L$ and $T=2\log_2 L$ for the same $p_m$. Increasing $T$ reduces the critical $p_u^{c,\mathcal{D}}$ from $0.514(12)$ to $0.3022(93)$, but the critical exponent remain unchanged within errorbars (from $\nu = 1.00(14)$ to $1.05(15)$).
Other data collapse results can be found in Appendix~\ref{app:supp_res}. 

Note that $\nu$ for the DT is different from the corresponding exponent for MIET -- not only is it much smaller, but it also potentially violates some theoretical bounds on critical parameters. The Chayes-Chayes-Fisher-Spencer (CCFS) bound~\cite{Chayes1986} is caused by the uncorrelated randomness in both space and time ($\nu \ge 2/(d+1) = 1$~\cite{Zabalo2023} with $d$ being the dimensionality of the quantum system), and is satisfied by MIET, while it could potentially be violated by the DT. On one hand, this could be because the locations of gates and measurements are known to the decoder (i.e.\@ the definition of ``random'' elements of the circuit is modified), on the other hand however, the errorbars on DT critical parameters are large, so we have not conclusively ruled out that $\nu$ is not above or equal to $1$. Other theoretical bounds inform whether the universality class of the transition changes under perturbations. The Harris criterion~\cite{Harris1974}, $\nu \ge 2/d = 2$, suggests that MIET is stable to quenched spatial disorder~\cite{Zabalo2023} (perfectly correlated in time), while DT is clearly not, and either vanishes (leaving the circuit not decodable) or changes its universality class. The less strict Harris-Luck bound~\cite{Luck1993}, $\nu \ge 1/d = 1$, informs that MIET is stable to quasiperiodic perturbations, while DT may or may not violate it, similarly to the CCFS bound.

Furthermore, we observe that the data collapses yielding a critical exponent $\nu$ much below $1$ tend to be worse, often requiring significantly larger system sizes to ascertain their correctness. If we discard these results, we find that the remaining estimates of $\nu$ are consistent within the errorbars across different circuit depths. This hints at the universality of the decodability transition, characterized by a fixed (non-running) critical exponent $\nu\approx 1$. This, interestingly, is very close to the prediction from the stochastic model we have developed in the previous section, where $\nu$ is independent of the details of the model, such as the log coefficient for the circuit depth. Moreover, mean-field universality classes often exhibit critical exponents that remain stable upon increasing the dimensionality of the problem. This motivates a deeper numerical investigation into whether the decodability transition exhibits similar dimensional insensitivity -- an avenue we explore in the next subsection.

Additionally, there is a question of the nature of the decodability transition, which appears to be second order, as evidenced by the scaling form underlying our data collapses. The decodability seems to be a function of a single parameter $(p-p_c) T^{1/\nu}$, where the circuit depth $T=a \log_2 L$. This logarithmic dependence on the system size evokes the scaling of superfluid density in the infinite-order Berezinskii-Kosterlitz-Thouless (BKT) transition, a universality class often describing the learnability transitions of global charge~\cite{Barratt2022field}; however, the scaling observed here involves $\ln L$, rather than $(\ln L)^2$ typically found in BKT-like scenarios. Likewise, the $\ln L$ dependence has a natural interpretation in the time dimension as the smallest important ``lengthscale'' in the system (the circuit depth). Therefore, taken together with our numerical analysis and the similarity to the stochastic model, we believe there is sufficient evidence for the decodability transition to be of second order and to exhibit universal behavior.

\subsection{Higher dimensional circuit models}

\begin{figure}[t]
    \centering
    \includegraphics[width=\columnwidth]{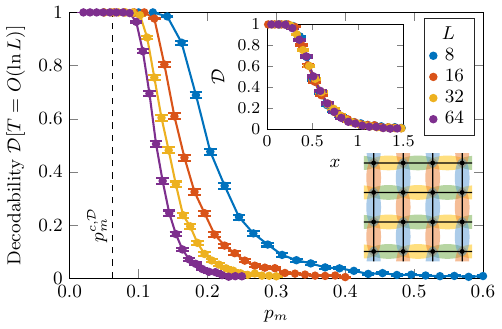}
    \caption{Decodability $\mathcal{D}[T]$ for a 2D circuit on a square lattice of $L\times L$ sites, for $p_u = 0.35$, and circuit depth $T = 2 \log_2 L = O(\ln L)$. The inset plot shows the data collapse for $p_m^{c,\mathcal{D}} = 0.062(27), \nu = 0.86(25)$. The diagram shows the pattern and order of unitary gate applications: green, yellow, blue, orange.}
    \label{fig:decodability_2D}
\end{figure}

When the considered circuit is run on a higher-dimensional lattice, we expect the results to mimic our findings in 1D, and thus to also lead to a decodability transition for logarithmic circuit depths. Note, for example, that the stochastic mean-field model of Eq.~\eqref{eq:PDFsm} is independent of the dimensionality and lattice structure, suggesting the same may be true for DT in the quantum circuit.

To confirm this assumption, we present numerics for a 2D square lattice circuit, where the unitary gates are applied in a sequence shown in the inset of Fig.~\ref{fig:decodability_2D}: green, yellow, blue, orange. The results for $p_u = 0.35$ and circuit depths of $T = 2 \log_2 L$ (where $L$ is the linear dimension) are shown in Fig.~\ref{fig:decodability_2D} and demonstrate the existence of the decodability transition. The data can be collapsed (see the inset plot) to yield the critical threshold of $p_m^{c,\mathcal{D}} = 0.062(27)$, and a critical exponent of $\nu = 0.86(25)$. Hence, we can see that in higher dimensions, DT is present for logarithmic circuit depths and is likely of second order.

Critical exponent $\nu$ is also consistent within the errorbar with the estimates from one dimension, leaving open the possibility that DT exhibits the same universality class across different lattice geometries. However, it is also plausible that the universality class is changed when increasing the dimensionality of the system, as is the case for many models. The number of unitary gates increases with the dimensionality, hence, it is conceivable that some geometries exhibit enough correlations to impact the universal properties of the transition. In that case, the randomness of the sign colors may become tightly related to the geometry of the lattice.

An interesting question is whether there exists a decodability transition for $T=O(L)$. Available numerical results rather show that the decodability is zero for any finite $p_m$ at large enough system sizes, and the random stochastic model also suggests that the transition should only occur for $T=O(\ln L)$. However, it is known that higher-dimensional models are more amendable to error correction~\cite{Kitaev2003, Fowler2012, IOlius2024, Behrends2024}, and the increased dimensionality often leads to quantum correlations having more impact on the dynamics. This could suggest that the stochastic model may not be able to capture all universal characteristics of the decodability transition in higher dimensions. Thus, it remains plausible that there may be a decodability transition at a very low $p_m$ in two- and higher-dimensional systems, akin to the results in Ref.~\cite{Li2023}, where the decodability transition was found at $T=O(\ln L)$ in one dimension and at $T=O(L)$ in two.

\subsection{Entangled initial states}
\label{sec:init_states}

\begin{figure}
    \centering
    \includegraphics[width=\columnwidth]{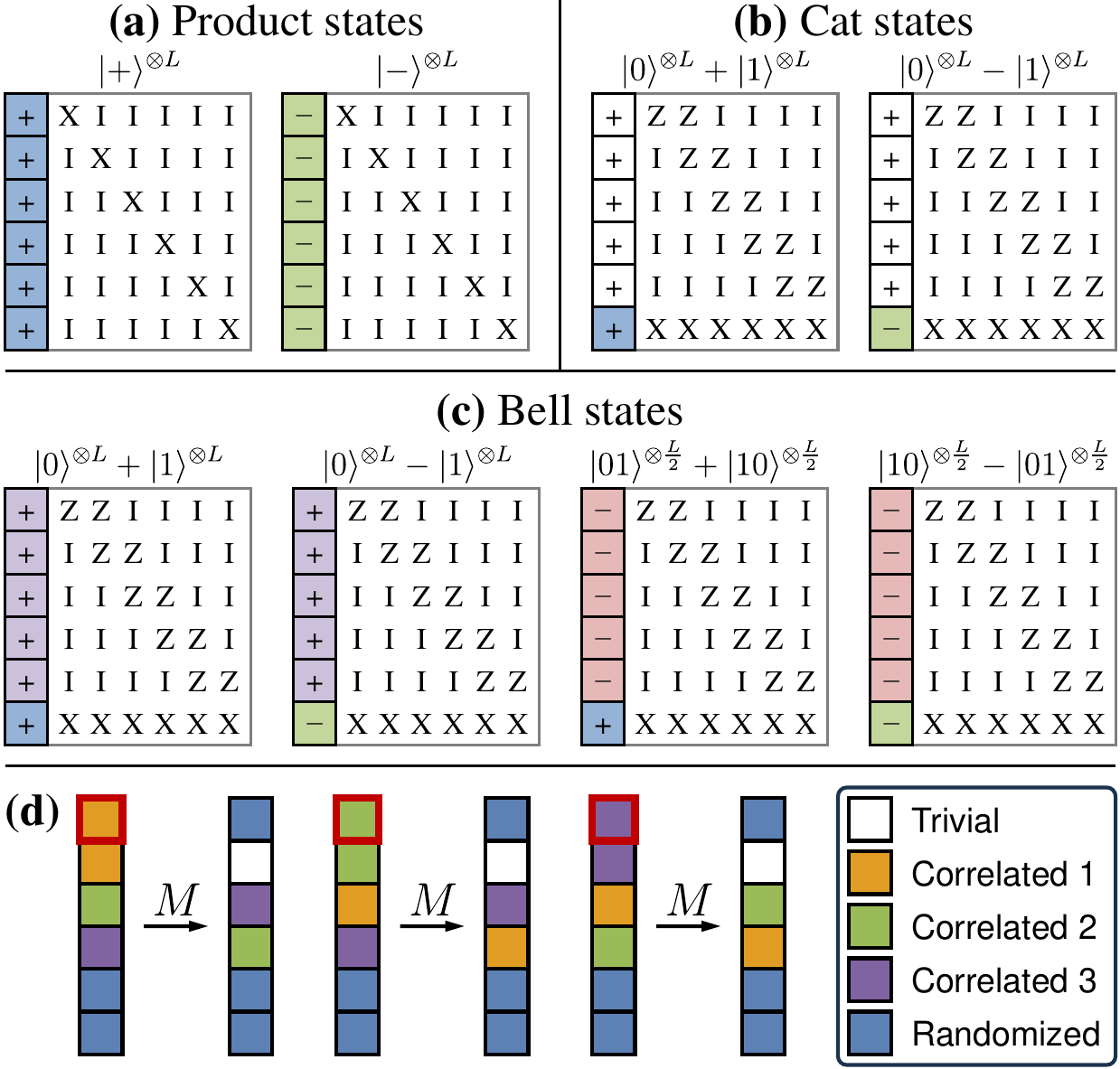}
    \caption{Examples of different initial states in the form of a tableau for $L=6$: (a)~product states, (b)~cat states, (c)~Bell states. Normalization is omitted. The signs correlated with the initial state are colored. (d)~Rules for sign color changes for the Bell states (cf.\@ Fig.~\ref{fig:signs_explanation}). There are three different correlated sign colors.}
    \label{fig:initial_states}
\end{figure}

The initial information so far was encoded as one of two initial product states: the all-plus state and the all-minus state. The tableaux corresponding to these states share the same stabilizers, with the only difference being the signs [see Fig.~\ref{fig:initial_states}(a)]. Note that encoding the initial bit into a product state with repeated local stabilizers is akin to a repetition code~\cite{Peres1985, Nielsen2010}, where a state is encoded as a repeated wave function forming a logical state. One can easily extend our protocol to any pair of states that share a common tableau, excluding the signs. A particularly interesting example is a pair of long-range entangled cat states [see Fig.~\ref{fig:initial_states}(b)], demonstrating that the decoding protocol can support non-local entanglement. The initial bit of information is here encoded into the sign of the global symmetry $\prod_i X_i$ and would be stable to $X$-errors, but not to e.g.\@ $Y$-errors.

Furthermore, the number of classical bits encoded within the choice of the initial state can be extended by initializing the circuit in one of $2^n$ states, where $n$ is the number of classical bits of information. An example is shown in Fig.~\ref{fig:initial_states}(c), where we write down four Bell states ($n=2$), for which there are three types of correlations with the initial state: sign corresponding to $Z_i Z_{i+1}$ stabilizers, sign corresponding to the global symmetry of $\prod_i X_i$, and sign corresponding to the multiplication of both. The multiplication rules for these three correlated sign colors are shown in Fig.~\ref{fig:initial_states}(d), which can be directly compared with Fig.~\ref{fig:signs_explanation}. These color multiplication rules bring to mind classical color mixing. Note that at least two types of correlated colors are needed to recover the initial state, otherwise, the state is not fully decodable.

The Sign-Color decoder for a general $n$-bit information remains essentially unchanged, with the key aspect being the ordering of the sign colors: trivial, all correlated colors, randomized. The total number of distinct correlated colors is $(2^n-1)$. Successful decoding requires $n$ distinct colors that serve as the dynamic syndrome, and these colors need to form an independent set, i.e., none can be obtained by multiplying other colors from the set. The protocol can be efficiently implemented using an $n$-bit vector of integers representing the trivial and correlated sign colors, while any random stabilizers can be removed from the tableau after each measurement. As $n$ increases, the average window of time for successful decoding decreases, and so does the number of qubits that can be impacted by errors before the initial information is irretrievably lost. This mirrors the tradeoff in quantum error correction between the encoded number of qubits and code distance: increasing the number of encoded bits in the same Hilbert space using the same method necessarily reduces the code distance, impacting the robustness of error protection.

Henceforth, the decoding algorithm can be formulated for non-trivial initial states with long-range entanglement, provided the error locations are known. We address the challenge of decoding with unknown error positions in the following section. Beyond this, an open question remains whether a slight modification to this protocol would enable the encoding and retrieval of more complex quantum information, i.e.\@ non-stabilizer states; we leave this possibility for future research.

\section{Decoding with unknown error locations}
\label{sec:decoding_unlocated}

We now consider an algorithm where the measurement locations are unknown (UL) to the decoder, thus treating the measurements as realistic errors within the circuit. Conversely, the positions of unitary gates are known to the decoder and will be utilized to enhance its efficiency and to form the dynamic syndrome in this particular protocol for decoding.

The Sign-Color decoder proposed in this section operates by benchmarking the circuit without measurements $p_m=0$ for identifying stabilizer generators for measurement to determine the initial state, i.e., to reveal the syndrome. Specifically, for given positions of unitaries, the measurement-free circuit ($p_m=0$) is used to obtain a list of $L$ stabilizer generators, $\{S_i\}$, which are time-evolved $\{X_i\}$ stabilizers of the initial state. We then consider realizations of a circuit with identical locations of unitaries but different measurement positions for a fixed measurement probability ($p_m > 0$). At the end of each circuit realization, we simultaneously measure all $\{S_i\}$ (which is possible since all generators commute). The average of $\{\langle S_i\rangle\}$ over all random measurement realizations can be interpreted as a list of weights indicating the likelihood of a stabilizer generator to identify the initial state. This list of weights is now the \textit{dynamic syndrome} of the UL decoding protocol. Finally, we choose the stabilizer with the largest weight, and identify its weight as the measure of decodability $\mathcal{D}$.

It is worth noting that a more complex algorithm for identifying the stabilizers to be measured for decoding the initial state is possible; in principle, such algorithms could achieve higher decodability (in comparison, the above-mentioned decoder is perfect for the KL problem). Therefore, $\mathcal{D}$ reported here should be considered as a lower bound on the true decodability. Additionally, although the measurement of $ S_i $ can yield $+1$, $0$, or $-1$, the outcome $-1$ is extremely unlikely after averaging over all measurement realizations (the stabilizer has a low probability of being anticorrelated). Introducing measurements can only preserve the stabilizer ($\langle S_i \rangle=+1$) or drive it to become independent of the initial state ($\langle S_i \rangle=0$). This allows us to reinterpret this average as a weight or probability.

\begin{figure}[t]
    \centering
    \includegraphics[width=\columnwidth]{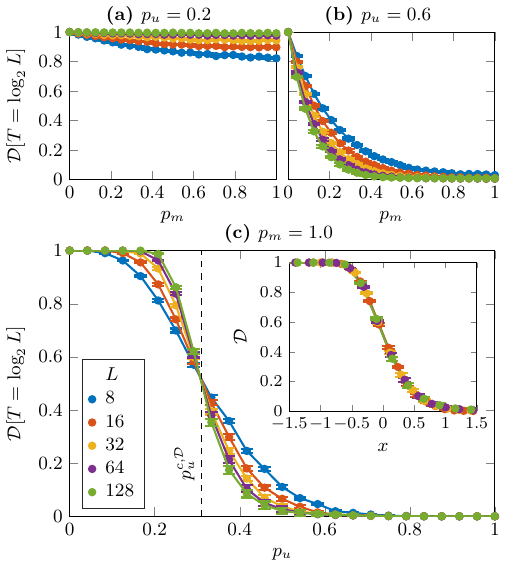}
    \caption{Decodability $\mathcal{D}$ when the location of measurements is unknown to the decoder, as a function of $p_m$ and $p_u$ for different system sizes $L$. (a)~$p_u = 0.2$, (b)~$p_u = 0.6$, (c)~$p_m = 1$. The legend in panel (c) applies in (a) and (b). The inset in (c) shows a data collapse with $p_u^{c,\mathcal{D}} = 0.309(17), \nu = 1.08(22)$, with $x=(p_u - p_u^{c,\mathcal{D}}) (\log_2 L)^{1/\nu}$.}
    \label{fig:unlocated_decodability}
\end{figure}

Hence, the decodability $\mathcal{D}$ is defined as:
\begin{equation}
    \mathcal{D} = \avg_U \max_i |\avg_M \langle S_i \rangle |,
\end{equation}
where the absolute value is taken to ensure that $\mathcal{D}\in[0,1]$ without loss of generality, and the averages are over configurations of unitary gates $U$ and measurements $M$. We present the results for circuit depths $T=\log_2 L$ in Fig.~\ref{fig:unlocated_decodability}. For $p_u=0.2$ and $0.6$ (panels a and b), we observe that $\mathcal{D}$ does not seem to present any transition as $p_m$ is increased; it only decreases steadily in both cases. However, as the system size is increased, we observe that $\mathcal{D}$ has a drastically different behavior for those two values of $p_u$. For $p_u=0.2$, the decodability tends to $1$ for all values of $p_m$, while it tends to $0$ when $p_u=0.6$. Therefore, the decodability must go through a transition as $p_u$ is varied. Fig.~\ref{fig:unlocated_decodability}(c) demonstrates this transition for the worst-case scenario, that is, when $p_m=1$ and errors happen everywhere.

The inset in Fig.~\ref{fig:unlocated_decodability}(c) shows the corresponding data collapse with the transition point $p_u^{c,\mathcal{D}} = 0.309(17)$, and a critical exponent of $\nu = 1.08(22)$. This value of $\nu$ is close to the exponent predicted by the stochastic mean-field model from Sec.~\ref{sec:simple_random}, while also being relatively close to the exponents extracted in the KL problem (Sec.~\ref{sec:emergence_of_decodability}, also see Appendix~\ref{app:supp_res} for a full table of extracted exponents). This suggests the universality of this (second-order) transition may match that of KL (and perhaps also the random model) and be independent of the specifics (such as the precise form of the logarithmic scaling of circuit depth or a specific cut through the phase diagram). Similar comments for the bounds for $\nu$ apply here as for the KL problem.

Interestingly, our results indicate that the transition point $p_u^{c,\mathcal{D}}$ is independent of the error frequency $p_m$: for three values of $p_m\in\{0.1,0.5,1.0\}$, we obtain $p_u^{c,\mathcal{D}} = 0.312(8), 0.308(8),$ and $0.306(16)$, respectively. This phenomenon can be explained by noticing that when $p_u<1/2$ and the circuit depth is logarithmic in the system size, there is a constant number of sites not acted upon by any unitary gate $U$ (at $p_u=1/2$ there is a scenario when all unitary gates are applied in a single bottom/top layer of the brickwork pattern). These untouched sites correspond to stabilizer generators that are not disturbed by any of the measurements (as they commute with the measurements), making it always possible to decode the initial state.

In fact, we can use this idea to approximate the transition point for $p_m=1$ by treating each site independently (ignoring correlations induced by the unitaries). In each time step, there is a $(1-p_u)^2$ probability that the unitary gate does not act on this site, which leads to the number of unaffected sites in circuit depth $T$ to be $(1-p_u)^{2T} L$. Transition to a non-decodable regime occurs when the number of unaffected sites dips below one, which corresponds to the critical point $p_u^{c,\mathcal{D}} = 1-2^{-1/(2a)}$, where $a = T/\log_2 L$. For $a=1$, this threshold corresponds to $p_u^{c,\mathcal{D}} = 1- 1/\sqrt{2} \approx 0.293$, which is surprisingly close to the numerical estimate, showing that the correlations can be safely ignored in this limit.

Our discussions on the UL decoding extend effectively to higher-dimensional circuits and other lattice geometries, much like the KL case. Consequently, we anticipate the presence of the UL decodability transition in other lattice structures, with the critical exponents for the transition being potentially close to the one-dimensional case. 

However, extending the UL decoding to other initial states or to different types of errors remains non-trivial. The intuition behind the UL problem being decodable is that the measurements of the Pauli $X$ do not change the correlated initial state stabilizers unless there exist intermediate unitary gates. In other words, the initial state is stabilized by the measurement operators, which leads to a transition even in the limit of $p_m=1$. For example, this is no longer true if the initial state is chosen from the set of states described in Fig.~\ref{fig:initial_states}(c), or for other types of errors, such as measurement of Pauli $Z$, or single-qubit dephasing gates. In these scenarios, we could expect that the system is only decodable near the $(p_u=0, p_m=0)$ point, and there is no longer an extended region of decodability for all values of $p_m$. These limitations stem from our choice of the decoding protocol, but this does not exclude the existence of sophisticated protocols with greater robustness against the incompatibility between the initial state and the errors.

\begin{figure}
    \centering
    \includegraphics[width=\columnwidth]{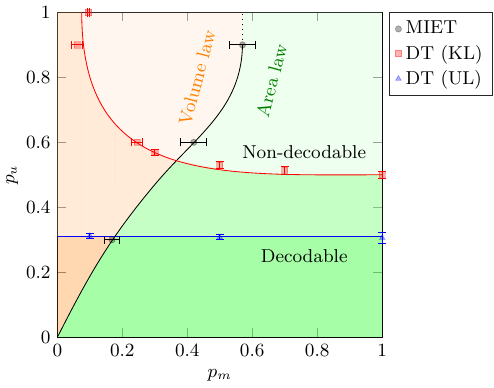}
    \caption{Phase diagram in the space of frequency of unitary gates $p_u$ and error rate $p_m$, showing the measurement-induced transition (MIET, black line) and two decodability transition lines: for known error locations (KL, red line), and for unknown locations (UL, red). We fix the circuit depth to $T=\log_2 L$. Different shading colors mark phases: volume law in orange, area law in green, while opacity signifies decodability (least opaque: non-decodable, medium opaque: decodable for KL and non-decodable for UL, most opaque: decodable for both KL and UL). Note the existence of the decodable volume-law phase for both KL and UL. The dotted line for MIET near $p_u=1$ shows that the transition is not well-defined in this limit.}
    \label{fig:1d_phase_diagram}
\end{figure}

To summarize, we can construct a decoder for the case where the error locations are unknown, which uses the knowledge of unitary gate locations to evaluate the dynamic syndrome, in the form of a list of weights for each stabilizer generator. We show numerically the existence of a decodable phase and the transition to a non-decodable phase, which appears to be independent of the error rate $p_m$. Our intuitive description is based on the unitaries being unable to disturb the stabilizers of the initial state on logarithmic time scales, while one can ignore the correlations generated by the unitary evolution. The decodable phase lies within the volume-law phase of the MIET, resulting in a region of decodable highly-entangled states in the phase diagram (see Fig.~\ref{fig:1d_phase_diagram}).

Practically, this establishes the existence of an encoder capable of storing information in complex many-body quantum states, even in the presence of noise, provided the error frequency remains below the threshold set by the MIET. The volume-law entanglement ensures that the information is inaccessible to third parties, in contrast to area-law states, where the information-carrying stabilizers are relatively easier to identify. Notably, the decoder succeeds even when the encoding process is noisy, requiring the knowledge of only the error types, not their locations. This demonstrates the possibility of reliable encoding and decoding of information using highly entangled states despite imperfect encoding dynamics.

\section{Discussion}
\label{sec:discussion}

In this work, we have presented a monitored stochastic quantum circuit that possesses simultaneously a decodability transition and a measurement-induced entanglement transition and hosts a decodable volume-law phase. We introduced the Sign-Color decoder, which tracks the dynamical syndrome to reveal the initial state, and is successful at efficiently decoding the state with either known or unknown error locations. Our results can be synthesized into a phase diagram (see Fig.~\ref{fig:1d_phase_diagram}) showing a decodable phase lying within the volume law phase of the associated MIET, allowing for complex encoding of information. Remarkably, by mapping the dynamics of stabilizer generators to a stochastic mean-field model, we have proposed that the decodability transition is a second-order phase transition, with critical exponents that are independent of certain details of the model, such as the precise circuit depth or lattice geometry. These aspects of our results provide solid evidence for universality applicable to a large class of systems.

Our research opens up numerous promising directions for future exploration. Our demonstration of encoding and decoding into an entangled initial quantum state, despite the dynamics being capable of generating a volume-law entangled state, provides a framework for utilizing more complex entangled states for quantum information processing. Efficiently controlling complex many-body states could prove useful for quantum cryptography, where the information is rendered inaccessible by a third party through scrambling. Furthermore, monitored quantum many-body systems often provide a useful statistical physics perspective for quantum error correction~\cite{Yang2025}, which in our case is manifested in the statistical properties of the dynamic syndrome of the decoder. In this context, the decoder could also incorporate mid-circuit feedback gates, which often leads to an absorbing subspace of states~\cite{Roy2020, Muller2022, Iadecola2023, Buchhold2022, Sierant2023, ODea2024, Sierant2023pol, Friedman2023, Piroli2023, Ravindranath2023, LeMaire2024, Qian2024}, potentially speeding up the decoding algorithms. 
Furthermore, improvements to the decoder for unknown locations of errors could be achieved by refining the choice of the stabilizer weights using neural networks, assisting the search for the optimal decoder. Our discovery of a decodable volume-law phase underscores potential applications for quantum computing, allowing for scrambling-enriched encoders, and in the future, for quantum information processing beyond break-even.

\begin{acknowledgments}
    The authors would like to thank Roberto Bondesan for useful discussions and Grace Sommers for detailed comments on the manuscript. D.~P.\@ is supported by the Engineering and Physical Sciences Research Council [grant number EP/S021582/1]. A.~P.\@ and M.~S.\@ were funded by the European Research Council (ERC) under the European Union's Horizon 2020 research and innovation programme (Grant Agreement No.\@ 853368). M.~S.\@ was also supported by the Engineering and Physical Sciences Research Council grant on Robust and Reliable Quantum Computing (RoaRQ), Investigation 004 [grant reference EP/W032635/1]. The authors acknowledge the use of the UCL Myriad High Performance Computing Facility (Myriad@UCL) and of the High Performance Computing cluster in the London Centre for Nanotechnology, and associated support services, in the completion of this work.
\end{acknowledgments}

\appendix

\section{Decodability transition in a stochastic model}
\label{app:simple_random}

In this appendix, we show that the decodability transition in the stochastic model of Eq.~\eqref{eq:PDFsm} exists only for logarithmic circuit depths $T \sim \ln L$ and show its scaling properties.

First, we note that the decodability can be derived from the probability density function of Eq.~\eqref{eq:PDFsm},
\begin{equation}
    \mathcal{D}[T] = 1 - \sum_{t=0}^T f_\text{os}(t) = 1-\left(1-(1-\mathcal{P})^{T+1}\right)^L.
\end{equation}
If we now assume logarithmic circuit depths $T = a \log_2 L$, $\mathcal{D}[T]$ exhibits a phase transition at the critical point $\mathcal{P}_c = 1 - 2^{-1/a}$, below which $\mathcal{D}[T] \to 1$, while above $\mathcal{D}[T] \to 0$ when $L \to \infty$. The finite-size scaling form near the transition region can also be simply derived. Note that the smallest lengthscale in this circuit is $T$. We find that if we use a rescaled variable $x = (\mathcal{P} - \mathcal{P}_c) T^{1/\nu}$ with $\nu = 1$, the decodability tends to a universal scaling function,
\begin{align}
    \mathcal{D}[T] &= 1-\left(1-\left(2^{-1/a}-\frac{x}{a \log_2 L}\right)^{a \log_2 L + 1} \right)^L\\
    &\to 1-\exp \left({-2^{-1/a} e^{-2^{1/a} x}} \right) = G[x].
\end{align}
The comparison to the usual finite-size scaling ansatz shows that this decodability transition has a critical exponent $\nu = 1$ independent of the coefficient of the log, $a$.

\section{Details of the finite size scaling collapse}
\label{app:fss}

The data collapse is performed using the following cost function $\varepsilon(p^c, \nu)$, which measures how the data collapses onto one curve. Here, we give an illustrative example of how to collapse the data for a measurement-induced transition, which can be easily extended to the decodability transition.

\begin{figure}[tb]
    \centering
    \includegraphics[width=\columnwidth]{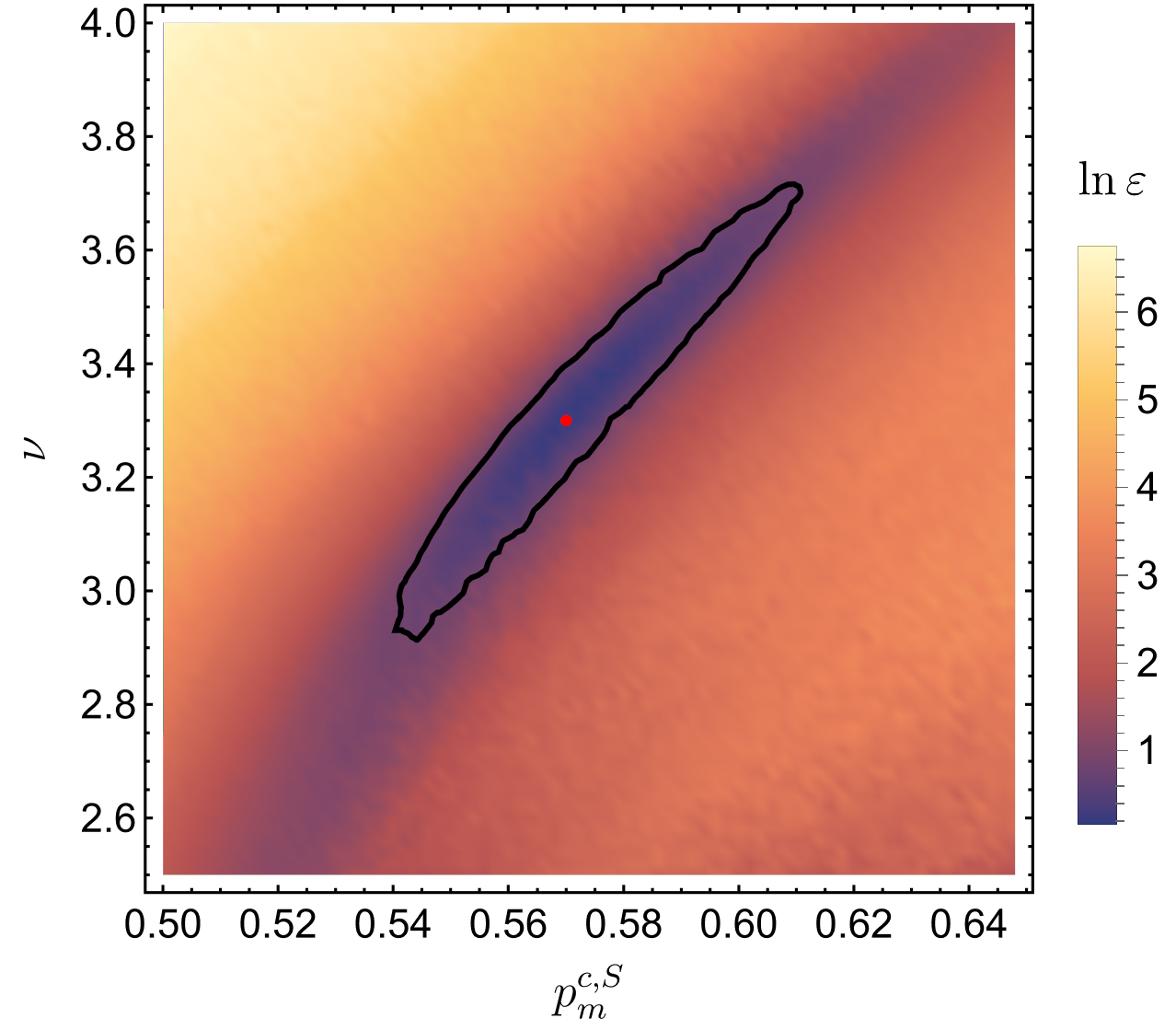}
    \caption{Cost function $\varepsilon(p_m^{c,S},\nu)$ for the measurement-induced entanglement transition at $p_u = 0.9$. The minimum is marked by a red dot, while the black region encompasses all values below $2\varepsilon_\text{min}$ and is used to estimate the error in the critical parameters.}
    \label{fig:cost_function}
\end{figure}

For each data point, we calculate a triple $(x,y,d)$ as follows: $x = (p_m - p_m^c) L^{1/\nu}$, $y = I_3$, and $d$ is the standard error in $I_3$. The triples are then sorted in increasing order by $x$. The cost function is then defined as
\begin{equation}
    \varepsilon(p^c, \nu) = \frac{1}{n-2} \sum_{i=2}^{n-1} \frac{(y_i - \bar{y}_i)^2}{\Delta_i^2},
\end{equation}
where
\begin{align}
    \bar{y}_i &= \frac{(x_{i+1}-x_i)y_{i-1} - (x_{i-1}-x_i)y_{i+1}}{x_{i+1}-x_{i-1}},\\
    \Delta_i^2 &= d_i^2 + \Big(\frac{x_{i+1}-x_i}{x_{i+1}-x_{i-1}}\Big)^2 d_{i-1}^2
    + \Big(\frac{x_{i-1}-x_i}{x_{i+1}-x_{i-1}}\Big)^2 d_{i+1}^2.
\end{align}
The value of $\varepsilon(p_m^{c, S}, \nu)$ is then averaged over different windows of $x$ near the transition region ($x \approx 0$). In the case of the MIET for $p_u = 0.9$ shown in Fig.~\ref{fig:1d_miet}, we use $x_\text{min} \in \mathcal{N}(-2.2,0.3), x_\text{max} \in \mathcal{N}(0.1,0.3)$, where $\mathcal{N}(\mu,\sigma)$ is the normal distribution with mean $\mu$ and standard deviation $\sigma$. We then find the minimum of $\varepsilon(p_m^{c, S}, \nu)$, which should be near 1 if the collapse is of good quality. The errors in $p_m^{c, S}$ and $\nu$ are estimated by drawing a region of $\varepsilon = 2 \varepsilon_\text{min}$, as shown in Fig.~\ref{fig:cost_function}.

\section{Supporting results}
\label{app:supp_res}

\begin{figure}
    \centering
    \includegraphics[width=\columnwidth]{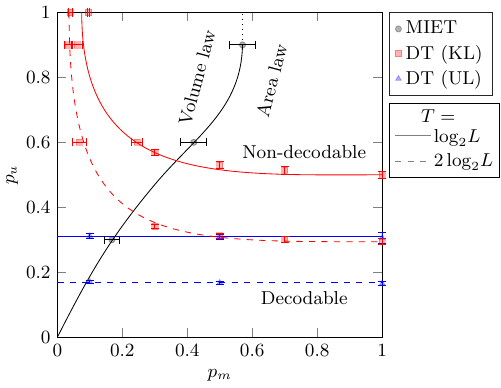}
    \caption{Phase diagram in the space of $p_u$ and $p_m$ (cf.\@ Fig.~\ref{fig:1d_phase_diagram}), showing measurement-induced entanglement transition (MIET), and the decodability transition (DT) for known error locations (KL) and unknown error locations (UL). We show results for circuit depths $T = \log_2 L$ (solid lines for DT) and $T = 2 \log_2 L$ (dashed lines for DT). \label{fig:1d_phase_diagram_complete}}
\end{figure}

This section presents all the results from data collapses performed in this work. Table~\ref{tab:supp_data} shows the values of extracted transition points and critical exponents of the correlation length $\nu$. The $p_u$ or $p_m$ that is reported with an error bar is the extracted transition point, while we fix the other parameter. Comparison between results for two different log coefficients of the circuit depth is shown in Fig.~\ref{fig:1d_phase_diagram_complete}.

\begin{table}[ht]
  \begin{tabular}{llll}
    \hline \hline
    Transition & $p_u$ & $p_m$ & $\nu$\\
    \hline
    MIET & 0.9 & 0.57(4) & 3.3(4)\\
    & 0.6 & 0.42(4) & 3.0(8)\\
    & 0.3 & 0.168(24) & 3.1(10)\\
    \hline
    DT (KL, $T = \log_2 L$) & 1.0 & 0.096(4) & 0.68(12)\\
    & 0.9 & 0.061(17) & 0.74(20)\\
    & 0.6 & 0.246(17) & 1.09(38)\\
    & 0.569(9) & 0.3 & 0.72(10)\\
    & 0.53(1) & 0.5 & 0.89(12)\\
    & 0.514(12) & 0.7 & 1.00(14)\\
    & 0.50(1) & 1.0 & 1.01(11)\\
    \hline
    DT (KL, $T = 2 \log_2 L$) & 1.0 & 0.040(5) & 0.49(20)\\
    & 0.9 & 0.034(11) & 0.49(27)\\
    & 0.6 & 0.067(21) & 0.80(48) \\
    & 0.341(6) & 0.3 & 0.94(12) \\
    & 0.3127(76) & 0.5 & 1.00(12) \\
    & 0.3022(93) & 0.7 & 1.05(15) \\
    & 0.294(9) & 1.0 & 1.06(13)\\
    \hline
    DT (KL, $T = 3 \log_2 L$) & 0.9 & 0.013(12) & 0.7(4)\\
    \hline
    DT (UL, $T = \log_2 L$) & 0.312(8) & 0.1 & 1.08(6)\\
    & 0.308(8) & 0.5 & 1.10(7)\\
    & 0.306(16) & 1.0 & 1.08(17)\\
    \hline
    DT (UL, $T = 2 \log_2 L$) & 0.171(5) & 0.1 & 1.05(10) \\
    & 0.168(4) & 0.5 & 1.08(8) \\
    & 0.166(6) & 1.0 & 1.08(13) \\
    \hline \hline
  \end{tabular}
  \caption{Data collapse results for measurement-induced entanglement transitions (MIET), and decodability transitions (DT) with known and unknown locations of errors (KL and UL) for specific circuit depths $L$.
  \label{tab:supp_data}}
\end{table}

\clearpage

\bibliography{refs}

\end{document}